# Exploring the Effects of Generative AI Assistance on Writing Self-efficacy

Yejoon Song[1], Bandi Kim[2], Yeju Kwon[2], Sung Park[3]*

[1]School of Business Innovation
[2]School of Humanities and Social Science
[3]School of Data Science and Artificial Intelligence
TAEJAE University, Seoul, Republic of Korea

[1]gabby7110@taejae.ac.kr
[3]*sjp@taejae.ac.kr

## 1 ABSTRACT

Generative AI (GenAI) is increasingly used in academic writing, yet its effects on students' writing self-efficacy remain contingent on how assistance is configured. This pilot study investigates how ideation-level, sentence-level, full-process, and no AI support differentially shape undergraduate writers' self-efficacy using a 2 × 2 experimental design with Korean undergraduates completing argumentative writing tasks. Exploratory findings indicate that AI assistance does not uniformly enhance self-efficacy: full AI support produced high but stable self-efficacy alongside signs of reduced ownership, sentence-level AI support led to consistent self-efficacy decline, and ideation-level AI support was associated with both high self-efficacy and positive longitudinal change. These patterns suggest that the locus of AI intervention, rather than the amount of assistance, is critical in fostering writing self-efficacy while preserving learner agency, establishing a foundation for future large-scale research.

## 2 INTRODUCTION

Argumentative writing is a fundamental yet challenging academic genre for university students, as it requires the integration of knowledge, reorganization of information, and persuasive articulation of original arguments [1]. Prior research has consistently reported underdeveloped argumentation skills among undergraduates, highlighting the need for diverse instructional strategies to enhance academic writing proficiency [2].

Recently, students have increasingly integrated GenAI tools such as ChatGPT into their academic writing processes. Studies indicate that these tools can improve writing efficiency and performance while also influencing students' writing self-efficacy and confidence [3, 4]. In higher education, writing self-efficacy is shaped by performance expectations, technological environments, and cognitive engagement [5, 6]. Given the substantial cognitive demands of academic writing, GenAI may provide meaningful support by compensating for underdeveloped essay-writing skills [2]. Understanding GenAI's influence on students' confidence and engagement is therefore critical for fostering autonomous and meaningful learning.

Despite growing acceptance of GenAI as a writing support tool, concerns persist regarding students' overreliance on these technologies. Although GenAI can reduce cognitive load, excessive dependence may undermine critical thinking, cognitive engagement, and students' sense of ownership over their work [7, 8]. Accordingly, there is a growing need for pedagogical frameworks that promote the responsible integration of GenAI by balancing its benefits with potential risks.

While prior studies have examined students' use of GenAI in writing tasks [7, 9] and noted its fragmented integration in educational contexts [10, 11], limited research has explored how different types of AI assistance balance benefits against risks. Addressing this gap, this pilot study investigates the differential effects of AI assistance on students' writing self-efficacy and evaluates the feasibility of the proposed experimental design for future large-scale research, with the goal of informing balanced and pedagogically sound GenAI integration in higher education.

## 3 RELATED WORK

Prior work [7] examined how heavy use of GenAI tools such as ChatGPT can lead to cognitive debt, a condition in which users progressively offload complex cognitive tasks to AI systems. Using neurophysiological measures and essay-writing tasks, they reported that extensive AI reliance reduced cognitive engagement and hindered higher-order thinking. These findings raise concerns regarding the potential risks of diminished agency and metacognitive effort in academic contexts. This concern aligns with this research, underscoring the importance of examining how different levels of AI assistance affect students' self-efficacy and cognitive ownership.

[4] examined the effects of GenAI platforms on undergraduate students' writing self-efficacy and narrative intelligence. Compared to students using traditional digital storytelling tools, those using AI-assisted writing platforms (e.g., Sudowrite, Jasper) showed greater gains in writing confidence and organizational skills. However, AI use did not significantly enhance students' sense of creative identity, suggesting that while AI can support structural and technical aspects of writing, it may not fully foster originality or personal voice. These findings underscore both the benefits and limitations of AI assistance in academic writing.

[12] examined the roles of writing self-efficacy and self-regulated learning (SRL) strategies in writing outcomes among English as a foreign language (EFL) students. They found that stronger writing self-efficacy and greater use of strategies such as goal setting, planning, and reviewing were associated with higher writing performance. However, many students showed low confidence and limited strategy use, highlighting a gap between learners' needs and their current practices. These findings suggest that writing outcomes may depend not only on external support but also on internal motivation and self-regulatory skills.

## 4 METHODOLOGY

### 4.1 Participants

This study was approved by the Ministry of Health and Welfare–designated Public Institutional Review Board (IRB; Protocol No. P01-202510-01-005). A total of 21 Korean undergraduate students were recruited via voluntary response sampling. Of these, 8 were excluded due to non-attendance or failure to comply with procedural requirements (camera activation or task instructions), leaving 13 participants for analysis. Participants were restricted to sophomore- to senior-level undergraduates to ensure familiarity with university-level academic writing. No restrictions were placed on academic major or institutional affiliation to enhance generalizability. Given the high prevalence of GenAI use among Korean undergraduates [13], this population was considered suitable for examining AI-assisted academic writing. All participants provided written informed consent, were assured of confidentiality, and were informed of their right to withdraw at any time without penalty.

### 4.2 Experimental Design

Drawing on prior research showing that students mainly use chatbots for idea generation and text revision [14], AI assistance was provided at two stages of argumentative writing: idea generation and text construction.

The experiment employed a 2 × 2 between-subjects design, producing four conditions based on the presence or absence of AI assistance at each writing stage (Table 1).

Table 1: Experimental Design

| Writing Conditions | Text Constructed by Self | Text Constructed with AI |
|---|---|---|
| Idea Generated by Self | No AI | Sentence AI |
| Idea Generated with AI | Ideation AI | Full AI |

A pre-survey assessed participants' writing priorities (idea generation, text construction, or both) and informed stratified group assignment to minimize self-efficacy bias.

### 4.3 Procedure

Participants completed a 100-minute online experimental session via Google Meet, with cameras kept on throughout. Following an orientation on the assigned AI usage type, participants completed two experimental cycles separated by a short break, each consisting of an argumentative writing task and a post-task self-efficacy questionnaire.

### 4.4 Materials

ChatGPT was selected as the GenAI tool due to its high usage among Korean undergraduates [13], which was expected to reduce novelty effects and support naturalistic interaction.

To control for potential confounding effects of affective or evaluative feedback on writing self-efficacy, the AI was instructed to provide neutral, task-focused responses only. Output was restricted to functional assistance corresponding to each group's assigned usage type, with motivational or evaluative language explicitly excluded. Full system prompts are provided in Appendix A.3.

Participants wrote two argumentative essays, each within 30 minutes (1,000–1,500 Korean characters). To minimize topic-related variance, the essays addressed two general topics requiring no specialized background knowledge: (a) Does recording volunteer hours undermine the value of voluntary service? and (b) Does true liberty emerge through law, or only without it? Topic order was counterbalanced across subgroups within each experimental condition to control sequencing effects.

### 4.5 Measures

Writing self-efficacy was measured using a continuous numerical rating scale ranging from 0 to 100. This format allows for fine-grained assessment of self-perceived writing ability and has been shown to enhance sensitivity and predictive validity in self-efficacy research [15, 16, 17]. The post-survey questionnaire was adapted from the writing self-efficacy scale [18]. Items from the cognitive and affective subdomains were selected and slightly modified to align with the argumentative writing tasks and experimental conditions (Table 2).

Table 2: Post-survey Questionnaire

| **Subcomponents** | **Evaluation Item Descriptions** |
|---|---|
| **Writing Ability** | I was able to identify and correct grammatical errors in my writing. |
| | I was able to select vocabulary appropriate to the given context. |
| | I was able to construct a coherent paragraph centered on a single main idea. |

| Subcomponents | Evaluation Item Descriptions |
|---|---|
| | I was able to write in accordance with the given topic of the essay. |
| | I was able to use evidence effectively to support my claims. |
| | I was able to evaluate whether the elements of my writing supported my argument and overall flow. |
| **Management of the Writing Process** | I was able to complete the writing task within the allotted time. |
| | I was able to independently resolve difficulties encountered during the writing process. |
| | I was able to write in accordance with the specified requirements and criteria. |
| | I was able to use my own language and personal writing style in the essay. |
| **Writing Experience** | I found it enjoyable to express my ideas through writing. |
| | I was able to write without experiencing excessive stress or discomfort. |
| | I did not feel overwhelmed or lost when starting the writing task. |
| | I did not feel anxious about my writing ability. |
| | I am not afraid of receiving feedback on my writing. |

# 5 RESULT

As a pilot study, the primary goal was not hypothesis testing but pattern identification and design validation for future large-scale studies.

## 5.1 Descriptive Statistics

Across 13 participants (Sentence AI: N = 5, Ideation AI: N = 2, Full AI: N = 4, No AI: N = 2), mean writing self-efficacy averaged across both writing occasions varied substantially by condition: Full AI ($\overline{M}$ = 82.2, SD = 6.4), Ideation AI ($\overline{M}$ = 80.1, SD = 16.1), Sentence AI ($\overline{M}$ = 76.6, SD = 7.2), and No AI ($\overline{M}$ = 61.8, SD = 24.0). The No AI group scored approximately 15 points below the weakest AI condition, with notably wider variability.

## 5.2 Effect Size Patterns

Pairwise comparisons revealed three noteworthy patterns. The primary comparison, Full AI versus No AI, yielded a large effect (g = 1.23, 95% CI: [−0.60, 3.07]), with Full AI users reporting 20.4 points higher in mean self-efficacy. Full AI also outperformed Sentence AI with a medium effect (g = 0.73). Sentence AI similarly showed a large advantage over No AI (g = 0.99, difference = 14.8 points). Ideation AI and Sentence AI differed minimally (g = 0.31), suggesting comparable overall impacts despite differing functional focuses. All confidence intervals spanned zero, precluding definitive conclusions.

## 5.3 Condition-specific Trajectories of Writing Self-efficacy Change

**Error! Reference source not found.** illustrates changes in writing self-efficacy from Time 1 to Time 2 across subscales and overall trends for participants in each group. Overall, Ideation AI showed the steepest increase ($\Delta M$ = +11.1, SD($\Delta$) = 16.2), though with substantial individual variation, as one participant improved markedly while another remained relatively stable. The No AI condition showed only a slight increase ($\Delta M$ = +1.3, SD($\Delta$) = 2.4), with participants largely maintaining their overall levels. Full AI remained nearly stable overall ($\Delta M$ = +0.5, SD($\Delta$) = 6.7), exhibiting mixed individual trajectories. In contrast, Sentence AI was the only condition showing decline, with all five participants experiencing decreased self-efficacy ($\Delta M = -5.7$, SD($\Delta$) = 5.6), suggesting sentence-level feedback may inadvertently undermine confidence with repeated exposure. These divergent patterns suggest AI support type influences not only self-efficacy levels but also self-efficacy stability across writing episodes. The strong Time 1–Time 2 correlation (r = 0.77) indicates reasonable measurement consistency.

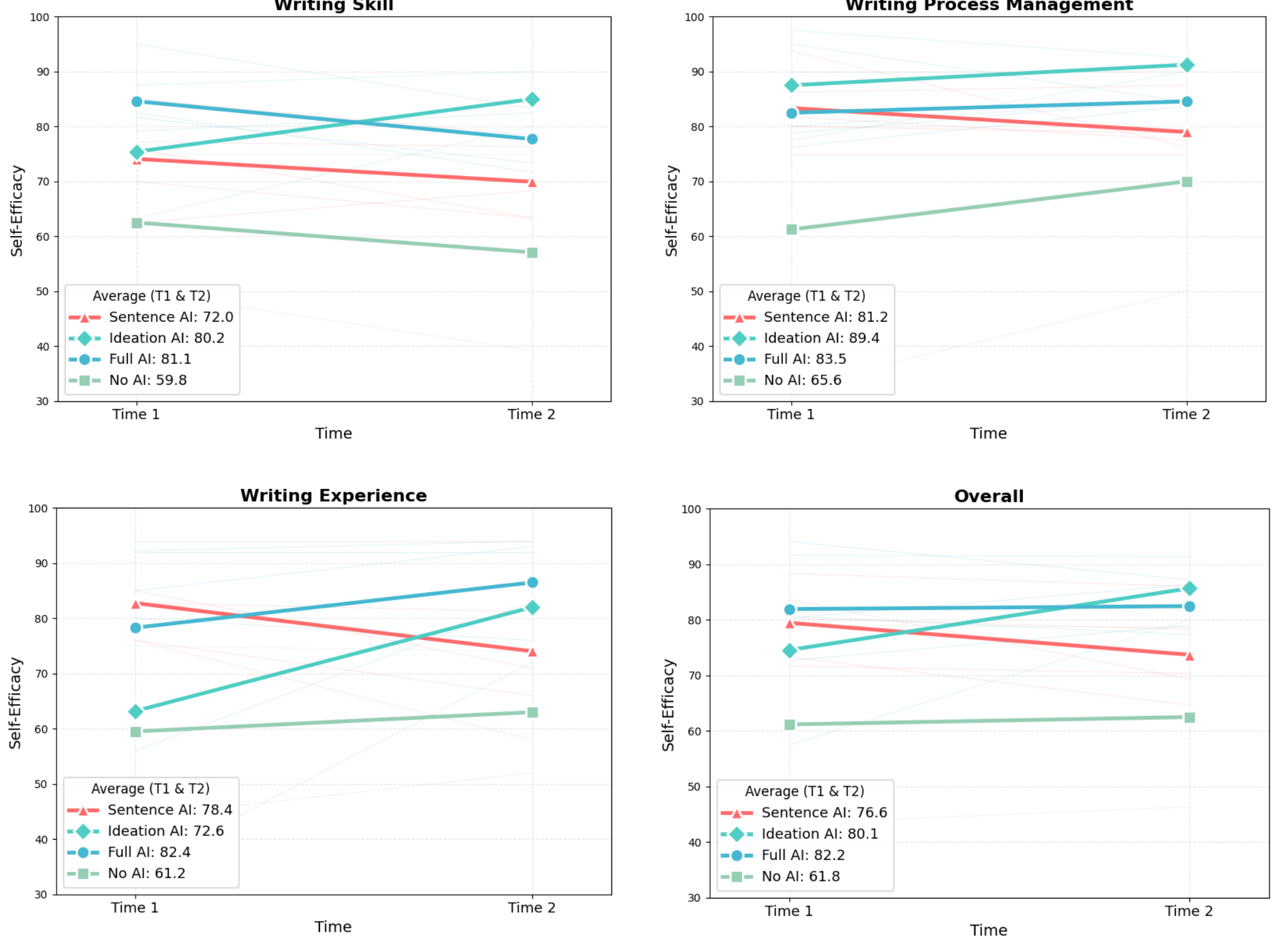

Figure 1: Writing Self-efficacy Over Time

### 5.4 Multidimensional Writing Self-efficacy

Analysis across three dimensions revealed distinct patterns by condition. Full AI showed balanced scores across dimensions ($\overline{M}$ = 81.2–83.5), whereas Ideation AI peaked in Process Management ($\overline{M}$ = 89.4) with a wide dimensional range ($\overline{M}$ = 72.6–89.4). Sentence AI showed moderate but uneven profiles ($\overline{M}$ = 72.0–81.2), and No AI consistently scored lowest ($\overline{M}$ = 59.8–65.6). These patterns suggest comprehensive AI support provides broad benefits, while targeted support strengthens specific competencies.

### 5.5 Exploratory Qualitative Observations

To contextualize the quantitative findings, post-task reflections and ChatGPT interaction logs were briefly reviewed for descriptive patterns (not formally coded). In the Sentence AI condition, most participants submitted complete drafts for holistic revision rather than requesting feedback on specific sentences; interactions were largely unidirectional and focused on stylistic or structural correction. In the Ideation AI condition, participants reported writing "in their own language" and expressed satisfaction with independently completed work; logs reflected bidirectional exchanges centered on idea clarification prior to independent drafting. In the Full AI condition, several participants described the AI-generated text as disconnected from their personal voice and reported reduced attachment; interaction logs showed both iterative exchanges involving participant-generated arguments and repeated prompts for extended AI drafts with minimal personal modification. In the No AI condition, one participant with a history of habitual AI use reported substantial difficulty writing independently when AI support was unavailable. These observations are reported to complement the quantitative results.

### 5.6 Measurement Validation

The continuous 0–100 scale demonstrated strong psychometric properties. Reliability was good for the total scale ($\alpha$ = 0.890) and Writing Experience subscale ($\alpha$ = 0.801), acceptable for Writing Ability ($\alpha$ = 0.701), though marginal for Process Management ($\alpha$ = 0.658), which is not uncommon for short subscales in pilot samples. Scale sensitivity analyses indicated a wide observed score range (43.3–94.1), 25 unique values across 26 observations, and no floor or ceiling effects, confirming the scale's sensitivity advantage over traditional Likert formats.

## 6 DISCUSSION

With small samples (N = 13), absolute mean comparisons are influenced by baseline writing self-efficacy differences. Full AI maintained the highest writing self-efficacy ($\overline{M}$ = 82.2) alongside a largely stable trajectory, yet two critical findings complicate this apparent success. First, qualitative responses revealed participants experienced diminished ownership and authenticity. Second, the No AI condition exhibited markedly low mean self-efficacy ($\overline{M}$ = 61.8), driven primarily by one participant who reported particularly low scores and was accustomed to AI-supported writing. This divergence suggests that Full AI may build self-efficacy with AI rather than for independent writing.

Ideation AI presents a more promising profile for sustainable skill development. This condition combined high self-efficacy ($\overline{M}$ = 80.1) with substantial improvement ($\Delta M$ = +11.1), alongside qualitatively positive engagement patterns. By allowing writers to externalize idea generation before and during drafting, Ideation AI may reduce the cognitive load associated with coordinating content development and linguistic formulation—a burden identified in cognitive process models of writing [19]. In this configuration, AI support appears to ease planning demands while preserving responsibility for sentence construction.

Sentence AI exhibited a troubling pattern: moderate overall self-efficacy ($\overline{M}$ = 76.6) obscured universal decline (5/5 participants, $\Delta M = -5.7$). Rather than functioning as a scaffold, sentence-level assistance may have operated as a comparative standard when delivered after draft completion. Consistent with social comparison theory [20], perceiving AI output as a superior benchmark may have triggered upward contrast, thereby reducing self-evaluations and contributing to the observed decline in writing self-efficacy.

These findings indicate that different AI configurations shape writing self-efficacy in different ways. When AI generates and revises text, self-efficacy may be tied to AI-supported performance rather than to the writer's own composing. Configurations that preserve the writer's responsibility for sentence construction may instead support more process-grounded self-efficacy.

## 7 LIMITATIONS

A key limitation of this pilot study is the small sample size, which constrained statistical power. This limitation stemmed from stringent inclusion criteria (i.e., Korean linguistic background, undergraduate status (sophomore to senior), and prior familiarity with GenAI) which enhanced internal precision but substantially restricted the eligible sample. Also, potential temporal effects warrant caution. Although Time 1–Time 2 trends differed across experimental groups, it remains unclear whether these differences are attributable to the manipulation or to familiarity, fatigue, or practice effects.

## 8 FUTURE DIRECTIONS

### 8.1 Rethinking the No AI Control Condition

Participants in the AI-supported conditions reported anxiety and uncertainty when AI assistance was reduced. These observations suggest that AI can no longer be treated as a simple controllable variable to be removed without consequence. Rather, future experimental designs should account for the residual and psychological effects of prior AI use, which may systematically influence behavior and writing self-efficacy across conditions.

In this context, the No AI condition warrants particular attention. Given that self-efficacy outcomes in the No AI condition varied substantially depending on participants' prior AI usage habits, future studies may benefit from explicitly measuring baseline AI reliance. One promising direction would be to stratify participants by their typical level of AI use.

### 8.2 Investigating the Full AI Paradox

AI-mediated writing may decouple writing self-efficacy from ownership—a historically co-varying relationship. Individuals may feel capable of producing acceptable outputs while experiencing reduced personal attachment or responsibility. Future research should examine how varying degrees of AI involvement reconfigure this relationship.

## 9 CONCLUSION

This pilot study reveals that GenAI assistance does not uniformly affect writing self-efficacy. Rather, outcomes depend critically on where AI intervenes in the writing process—not simply whether AI is used. This configuration-dependency represents a key methodological insight: even with the same AI tool, structural choices about when and how assistance is provided produced distinct outcomes. Ideation support fostered growth and preserved agency, sentence-level feedback led to decline, and comprehensive assistance produced stability alongside diminished ownership. Although the Ideation AI and No AI conditions had small sample size (N = 2), these divergent patterns underscore that design choices in AI-assisted writing systems fundamentally shape learner beliefs about their writing capabilities.

Equally important, this study validates a trajectory-based approach to measuring writing self-efficacy change. By tracking within-person change across two writing episodes, we identified critical temporal dynamics that aggregate-level comparisons would obscure. What seems adequate in the moment may mask erosion over time. This discrepancy highlights trajectory analysis as a more sensitive indicator of AI's long-term effects on learner beliefs.

Large-scale replication is needed to confirm these configuration-specific patterns, and longitudinal studies should examine whether observed trajectories persist beyond experimental contexts. By demonstrating that AI assistance configuration shapes both writing self-efficacy levels and developmental trajectories, this pilot study provides a foundation for designing AI writing support that enhances rather than undermines learner agency.

## A APPENDICES

### A.1 Session Structure

Each participant completed a 100-minute online session structured as follows:

1. Orientation (10 minutes)
2. Writing Task 1 (30 minutes)
3. Post-task self-efficacy survey (10 minutes)
4. Break (5 minutes)

5. Writing Task 2 (30 minutes)
6. Post-task self-efficacy survey (10 minutes)

### A.2 ChatGPT Version and Access

AI responses were generated using the publicly available free web-based version of ChatGPT (OpenAI). The experiments were conducted in November and December 2025. At the time of data collection, the specific underlying model identifier was not visible in the free-tier interface. All interactions were conducted via the standard ChatGPT web interface without API access, external plug-ins, or parameter customization.

### A.3 AI Configuration Prompts by Experimental Condition

In AI-supported conditions, participants initiated a new ChatGPT session prior to each writing task and inserted the assigned system prompt as the first message. The prompts below reflect the instructions administered during the experiment.

#### *A.3.1 Sentence-Level AI Condition*

Participants independently generated ideas and were permitted to use ChatGPT only for sentence-level drafting and revision.

> In all conversations, do not use praise, exclamations, rhetorical embellishments, or unnecessary elaboration.
> Respond concisely and accurately to the user's explicit request only.
> Do not introduce new ideas, expand the topic, provide suggestions, interpretations, examples, or background explanations beyond the scope of the request.
> For requests involving sentence editing or reorganization, focus strictly on grammatical and stylistic corrections.
> Adding new content or extending meaning is strictly prohibited.
> All responses must be limited to factual clarification or sentence-level revision.

#### *A.3.2 Ideation-Level AI Condition*

Participants were permitted to use ChatGPT exclusively during the idea generation stage and were required to compose all text independently.

> In all conversations, do not use praise, exclamations, rhetorical embellishments, or unnecessary elaboration.
> Respond concisely and accurately to the user's explicit request only.
> Interaction is restricted to idea generation.
> Develop perspectives, arguments, and conceptual approaches directly related to the topic.
> Do not draft complete sentences, revise written text, or modify expressions beyond the scope of idea exploration.

#### *A.3.3 Full-Process AI Condition*

Participants were permitted to use ChatGPT during both idea generation and sentence construction stages.

> In all conversations, do not use praise, exclamations, rhetorical embellishments, or unnecessary elaboration.
> Respond concisely and accurately to the user's explicit request only.
> Do not introduce information beyond the requested scope.
> All responses must remain strictly task-focused.

#### *A.3.4 No-AI Control Condition*

No system prompt was administered. Participants completed all writing tasks independently without generative AI assistance.